\documentclass{pasj01}
\usepackage{epstopdf}
\usepackage{color}
\Received{$\langle$2017 May 15$\rangle$}
\Accepted{$\langle$2017 June 2$\rangle$}
\Published{$\langle$publication date$\rangle$}

\begin{document}

\title{Pilot KaVA monitoring on the M87 jet: confirming the inner jet structure and superluminal motions at sub-pc scales}
\author{
Kazuhiro Hada,$^{1,2,*}$
Jong Ho Park,$^{3}$
Motoki Kino,$^{1,4}$
Kotaro Niinuma,$^{5}$
Bong Won Sohn,$^{4,6}$
Hyun Wook Ro,$^{7}$
Taehyun Jung,$^{4}$
Juan-Carlos Algaba,$^{4}$
Guang-Yao Zhao,$^{4}$
Sang-Sung Lee,$^{4,6}$
Kazunori Akiyama,$^{8}$
Sascha Trippe,$^{3}$
Kiyoaki Wajima,$^{4}$
Satoko Sawada-Satoh,$^{9}$
Fumie Tazaki,$^{1}$
Ilje Cho,$^{4,6}$
Jeffrey Hodgson,$^{4}$
Jeong Ae Lee,$^{4,6}$
Yoshiaki Hagiwara,$^{10}$
Mareki Honma,$^{1,2}$
Shoko Koyama,$^{11}$
Junghwan Oh,$^{3}$
Taeseak Lee,$^{3}$
Hyemin Yoo,$^{7}$
Noriyuki Kawaguchi,$^{12}$
Duk-Gyoo Roh,$^{4}$
Se-Jin Oh,$^{4}$
Jae-Hwan Yeom,$^{4}$
Dong-Kyu Jung,$^{4}$
Chungsik Oh,$^{4}$
Hyo-Ryoung Kim,$^{4}$
Ju-Yeon Hwang,$^{4}$
Do-Young Byun,$^{4}$
Se-Hyung Cho,$^{4}$
Hyun-Goo Kim,$^{4}$
Hideyuki Kobayashi,$^{1}$
Katsunori M. Shibata$^{1,2}$
}

\altaffiltext{1}{Mizusawa VLBI Observatory, National Astronomical Observatory of Japan, 2-21-1 Osawa, Mitaka, Tokyo 181-8588, Japan}
\altaffiltext{2}{Department of Astronomical Science, The Graduate University for Advanced Studies (SOKENDAI), 2-21-1 Osawa, Mitaka, Tokyo 181-8588, Japan}
\altaffiltext{3}{Department of Physics and Astronomy, Seoul National University, Gwanak-gu, Seoul 08826, Republic of Korea}
\altaffiltext{4}{Korea Astronomy and Space Science Institute, Yuseong-gu, Daejeon 34055, Korea} 
\altaffiltext{5}{Graduate School of Sciences and Technology for Innovation, Yamaguchi University, Yoshida 1677-1, Yamaguchi, Yamaguchi 753-8512, Japan}
\altaffiltext{6}{Department of Astronomy \& Space Science, University of Science \& Technology, 217 Gajeong-ro, Daejeon, Republic of Korea}
\altaffiltext{7}{Department of Astronomy, Yonsei University, 134 Shinchondong, Seodaemungu, Seoul 120-749, Republic of Korea}
\altaffiltext{8}{Massachusetts Institute of Technology, Haystack Observatory, 99 Millstone Road, Westford, MA 01886, USA}
\altaffiltext{9}{Graduate School of Science and Engineering, Kagoshima University, 1-21-35 Korimoto, Kagoshima-shi, Kagoshima 890-0065, Japan}
\altaffiltext{10}{Toyo University, 5-28-20 Hakusan, Bunkyo-ku, Tokyo 112-8606, Japan}
\altaffiltext{11}{Max-Planck-Institut f\"ur Radioastronomie, Auf dem H\"ugel 69, Bonn, 53121, Germany}
\altaffiltext{12}{Shanghai Astronomical Observatory, Chinese Academy of Sciences 80 Nandan Road, Xuhui District, Shanghai 200030, China}

\email{kazuhiro.hada@nao.ac.jp}


\maketitle

\begin{abstract}
We report the initial results of our high-cadence monitoring program on the radio jet in the active galaxy M87, obtained by the KVN and VERA Array (KaVA) at 22\,GHz. This is a pilot study that preceded a larger KaVA-M87 monitoring program, which is currently ongoing. The pilot monitoring was mostly performed every two to three weeks from December 2013 to June 2014, at a recording rate of 1\,Gbps, obtaining the data for a total of 10 epochs. We successfully obtained a sequence of good quality radio maps that revealed the rich structure of this jet from $\lesssim$1\,mas to 20\,mas, corresponding to physical scales (projected) of $\sim$0.1--2\,pc (or $\sim$140--2800 Schwarzschild radii). We detected superluminal motions at these scales, together with a trend of gradual acceleration. The first evidence for such fast motions and acceleration near the jet base were obtained from recent VLBA studies at 43\,GHz, and the fact that very similar kinematics are seen at a different frequency and time with a different instrument suggests these properties are fundamental characteristics of this jet. 
This pilot program demonstrates that KaVA is a powerful VLBI array for studying the detailed structural evolution of the M87 jet and also other relativistic jets. 
\end{abstract}

\section{Introduction}

Very long baseline interferometry (VLBI) is a powerful technique for probing compact nonthermal radio sources and phenomena in the Universe. In particular, VLBI plays a major role in studying relativistic jets in active galactic nuclei (AGN), since the ultra-high-resolution capability allows direct imaging of the sites of e.g., collimation, acceleration and production of high-energy $\gamma$-ray emission in the jets, which generally take place at pc-to-subpc scales or even less. Over the past decades, tremendous observational efforts have been made with various VLBI arrays around the world (e.g., VLBA, EVN and LBA) to elucidate the relativistic jets. Nevertheless, a number of fundamental questions on jet physics still remain to be answered, requiring further imaging and monitoring of the jet-formation scales with high-resolution VLBI.  

The KVN and VERA Array (KaVA) has recently emerged as the first regularly operated international VLBI network in East Asia, consisting of three 21-meter dishes in Korea (KVN; e.g., \cite{lee2014}) and four 20-meter dishes in Japan (VERA; \cite{kobayashi2003}). Over the last years, we made efforts to combine the two arrays in collaboration among the Korea Astronomy and Space Science Institute (KASI), National Astronomical Observatory of Japan (NAOJ) and associated universities. The aim of this project is to form a single higher performance array by complementing each other. Indeed, the increases in the number of baselines (from 3/6 for KVN/VERA to 21 for KaVA) and the baseline coverage (from 300-400\,km/1000-2300\,km for KVN/VERA to 300-2300\,km for KaVA) significantly improve the overall array sensitivity and the imaging performance, as first demonstrated by \citet{matsumoto2014} and \citet{niinuma2014}. 

Following a commissioning phase lasting several years, partial (risk-shared) regular operation of KaVA began in 2014. The common frequency bands of 22\,GHz and 43\,GHz are available, and a data-recording rate of 1\,Gbps is currently supported. All data observed by KaVA are correlated at the Korea-Japan Joint VLBI Correlator installed at KASI (the so-called Daejeon hardware correlator; \cite{oh2012, lee2015a}). A feature of KaVA is that the array operates for a quasi-full year, except for the annual (6--8)-week maintenance period that starts from mid-June. This offers an ideal opportunity to monitor structural evolution of relativistic jets via detailed multi-epoch observations. 

While some early KaVA monitoring results on bright blazars have already been published~\citep{oh2015, an2016}, here we focus on the radio galaxy M87. M87 is one of the nearest AGN with a powerful relativistic jet. Due to its proximity~(16.7\,Mpc; \cite{blakeslee2009}) and the huge mass of the central black hole ((3--6)$\times 10^9M_{\odot}$; \cite{gebhardt2011}; \cite{walsh2013}), this source offers an exceptional opportunity to resolve the jet-formation scales with VLBI (e.g., \cite{junor1999}; \cite{hada2011}). Recent high-sensitivity VLBI observations at 86\,GHz revealed a limb-brightened, wide-opening-angle jet launching at a scale of $\lesssim$10\,$R_{\rm s}$~\citep{hada2016, kim2016}. Moreover, extensive analyses of multi-frequency VLBI images revealed a parabolic collimation over a distance from $z$$\sim$100\,$R_{\rm s}$ to $z$$\sim$$10^5$\,$R_{\rm s}$~\citep{asada2012, hada2013a}. These results suggest that magneto-hydrodynamic (MHD) processes play a key role in collimating the M87 jet~(e.g., \cite{nakamura2013}). 

Compared to the progress in morphology analysis, the structural evolution of this jet is still controversial. Early multi-epoch VLBI studies of M87 (at sparse intervals of months-years) often reported slow or quasi-stationary motions at the pc-to-subpc scales~\citep{reid1989, dodson2006, ly2007, kovalev2007}. However, such low speeds would be insufficient to explain the known large jet-to-counter-jet brightness ratio at the same scales (typically $>$10). To date, the most detailed program was performed by Walker et al. in 2007--2008 with VLBA at 43\,GHz~\citep{walker2008, walker2016}, where M87 was monitored every three weeks from 2007 Jan-Aug and every $\sim$five days from 2008 Jan-Apr. In contrast to the previous knowledge, they first noted the existence of fast ($\gtrsim$(1--2)\,$c$) motions near the jet base. Using the same data, the recent sophisticated image analysis by \citet{mertens2016} have further suggested the presence of flow acceleration at these scales.  Nevertheless, it is still unclear whether such fast motions were just a temporal event or not, and also whether similar motions can be seen at other frequencies or not.  

Therefore, there is a growing consensus on the necessity of a high-cadence (every a few weeks or less), continuous monitoring program to properly understand the kinematics of the M87 jet. Unfortunately, the VLBA's massive program is triggered only occasionally along with unpredictable TeV events, preventing a monitoring of the source on a regular basis. Since M87 is one of the few jet sources that gravitational scales of $\ll10^{3-4}\,R_{\rm s}$ are directly resolvable, having a regular monitoring program should be of great value for testing the magnetically driven jet paradigm by comparing with theories, as well as rapid follow-up of high-energy events. This is our primary motivation for launching a new monitoring program of this jet with KaVA. 

In this paper we report on the initial results from our first KaVA--M87 monitoring program. This is a pilot study conducted at the early stage of our KaVA operation, and thus the present paper primarily focuses on the validation of KaVA's imaging and monitoring performance on this jet. More detailed analyses (including newer data) and comparisons with some specific theoretical models are currently prepared in separate papers (Park et al. in prep; Ro et al. in prep.).  In the next section we describe our observations and data reduction. In Sections 3 and 4, the results on imaging and kinematics are presented. In the final section we will summarize the paper and describe our future prospects on this project. Throughout the paper we adopt $D=16.7\,{\rm Mpc}$ and $M_{\rm BH} = 6.0 \times 10^9\,M_{\odot}$, corresponding to $1\,{\rm mas} = 0.08\,{\rm pc} = 140\,R_{\rm s}$.

\begin{table*}[htbp]
 \begin{minipage}[t]{1.0\textwidth}
  \tbl{KaVA observations of M87}{
    \begin{tabular*}{1.0\textwidth}{@{\extracolsep{\fill}}lccccccc}
    \hline
    \hline
    UT Date  & Stations & On-source time & Beam size (natural) & Beam size (uniform) & $I_{\rm peak}$ & $I_{\rm rms}$ & $I_{\rm peak}/I_{\rm rms}$\\
             & & (min.) & (mas$\times$mas, deg.) & (mas$\times$mas, deg.) & $\left(\frac{\rm mJy}{\rm beam}\right)$ & $\left(\frac{\rm mJy}{\rm beam}\right)$ &  \\
             & (a) & (b) & (c) & (d) & (e) & (f) & (g)    \\
    \hline
    2013 Dec  5  & KaVA & 65 &$1.31\times 1.14, -18$ & $1.20\times 1.01, -27$ & 1661 & 0.94 & 1722\\
    2013 Dec 26  & KaVA & 65 &$1.61\times 1.23, -43$ & $1.32\times 1.06, -44$ &  1317 & 0.80 & 1638\\
    2014 Jan 15  & KaVA & 65 &$1.45\times 1.23, -41$ & $1.22\times 1.05, -42$ & 1277 & 0.74 & 1732\\
    2014 Mar  2  & KaVA & 270 &$1.46\times 1.21,  +3$ & $1.18\times 1.03,  +2$ & 1373 & 0.42 & 3306\\
    2014 Mar 15  & KaVA & 270 &$1.34\times 1.13, -16$ & $1.11\times 0.98, -17$ & 1400 & 0.39 & 3603\\
    2014 Apr  3  & KaVA & 270 &$1.32\times 1.22, +14$ & $1.10\times 1.02, +25$ & 1348 & 0.80 & 1676\\
    2014 Apr 16  & KaVA & 270 &$1.27\times 1.09, -18$ & $1.08\times 0.96, -22$ & 1384 & 0.35 & 3973\\
    2014 May  3  & KaVA & 270 & $1.47\times 1.26,+12$ & $1.15\times 1.02, +13$ & 1408 & 0.42 & 3390\\
    2014 Jun  2  & KaVA, $-$IRK & 270 & $1.08\times 0.91, -18$ & $1.00\times 0.84, -28$ & 1171 & 0.67 & 1748\\
    2014 Jun 14 & KaVA, $-$IRK & 270 & $1.34\times 1.10,-12$ & $1.10\times 0.97, -13$ & 1167 & 0.46 & 2562\\
    \hline
    \end{tabular*}}\label{tab:img_prm}
  \end{minipage}
   Notes: (a) Participating stations. KaVA indicates all seven KaVA stations. IRK represents the VERA-Iriki station. A minus sign before a station name means the absence of that station. (b) On-source time of M87. (c) Synthesized beam with a natural weighting scheme. (d) Synthesized beam with a uniform weighting scheme. (e) Peak intensity of M87 images with a natural weighting scheme. (f) Off-source rms image noise level of M87 images under a natural weighting scheme. (g) Dynamic range calculated with $I_{\rm peak}$ and $I_{\rm rms}$. 
\end{table*}

\section{Observations and Data Reduction}
From December 2013 to June 2014 M87 was densely monitored with KaVA at 22\,GHz. The period was in a transition phase between internal commissioning and regular science operation of KaVA, so the data used here are the combined results of these two stages. The monitoring was made at sampling intervals of two to three weeks. The observations were largely successful, but two particular sessions performed on 2016 February 3 and 25 were lost due to recording issues. In addition, the data taken on 2014 May 15 was poor in quality due to the loss of VERA-Ishigaki and severe weather conditions at VERA-Mizusawa, which prevented us from obtaining a reliable jet image. As summarized in Table~1, we eventually obtained our dataset for a total 10 epochs during the seven months. For the first three epochs (Dec/05, Dec/25, Jan/15), the on-source time of M87 was relatively short because the source was intermittently scanned in a multi-source survey program, while for the rest of the epochs M87 was observed with a quasi-continuous mode. Weather conditions were generally stable at the KVN and VERA-Mizusawa/Iriki stations, while the weather at the two southern stations VERA-Ishigaki/Ogasawara was a little more variable (e.g., rain showers). Left-hand circular polarization was received and sampled with a 2-bit quantization. At the VERA stations, a single-beam mode was used in KaVA observations. All the data were recorded at 1\,Gbps (256\,MHz bandwidth, 16\,MHz $\times$ 16 subbands) and correlated by the Daejeon correlator. 

The initial data calibration was performed using the National Radio Astronomy Observatory (NRAO) Astronomical Image Processing System (AIPS) based on the standard VLBI data reduction procedures. A priori amplitude calibration was applied using the measured system noise temperature and the elevation-gain curve of each antenna. We calibrated the amplitude part of bandpass characteristics at each station using the auto-correlation data. Following the amplitude calibration, fringe-fitting was performed to calibrate the visibility phases, and finally the data were averaged over each sub-band. Imaging and self-calibration were performed in the Difmap software~\citep{shepherd1994} in the usual manner. 

For KVN data correlated with the Daejeon correlator, there is a known ``amplitude-loss'' problem due to the accumulation of multiple losses through signal processing \citep{lee2015b}. For such KVN data \citet{lee2015b} specified an  amplitude correction factor of 1.35 (for the data correlated before 2015 March) or  1.30 (for the data correlated after 2015 March). As for KaVA data, the corresponding values are currently being examined (Oh et al. in prep), but we may reasonably assume the same correction factors, since the backend system of VERA is similar to that of KVN. Therefore in the present analysis of our KaVA data, we rescaled the amplitude of the calibrated visibilities by a factor of 1.35. In Section 3.2 we show that this factor is indeed reasonable.

\section{Results and Discussion}

\subsection{Images}

The first KaVA image of M87 at 22\,GHz was presented in \citet{niinuma2014}, where the observation was made at 128\,Mbps with a relatively short ($\sim$60\,min) integration time. While the early KaVA image nicely demonstrated the improvement of image quality over the only-VERA case, the image dynamic range was still limited to $\sim$1000 and the jet emission was barely detected down to $\lesssim$10\,mas from the core.   

\begin{figure}[ttt]
 \begin{center}
  \includegraphics[width=\columnwidth]{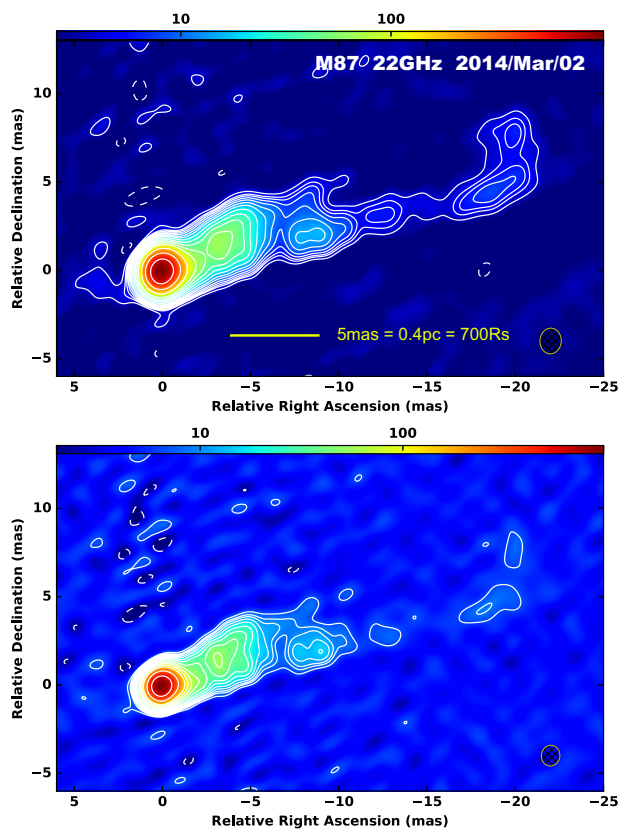}
 \end{center}
 \caption{Representative KaVA 22\,GHz images of the M87 jet, observed on 2014 March 2nd. (Top) naturally weighted image describing the overall detected jet emission. (Bottom) uniformly weighted image resolving the inner jet region. For both images the contours start from $-$1, 1, 2, $2^{1/2}$, 4.... times 3$\sigma$ ($1\sigma=0.39/0.64\,{\rm mJy\,beam^{-1}}$ for the top/bottom images, respectively).}\label{fig:m87map}
\end{figure}

To describe the typical imaging quality of our present 1\,Gbps KaVA dataset, we show two representative images of M87 in Fig.~\ref{fig:m87map}, which are the same data taken on 2014/Mar/2 but different uv-weighting and restoring beams are applied. 

The naturally weighted image (the top panel of Fig.~\ref{fig:m87map}) illustrates the overall structure of the M87 jet. Thanks to the improvement of our data-recording rate, KaVA detected the extended jet emission down to $\sim$20\,mas very clearly (at SNR much higher than 5\,$\sigma$). The resulting dynamic range of our KaVA 1\,Gbps images was 1600-4000, depending on the number of stations, weather condition and on-source time. 

\begin{figure}[ttt]
 \begin{center}
  \includegraphics[width=\columnwidth]{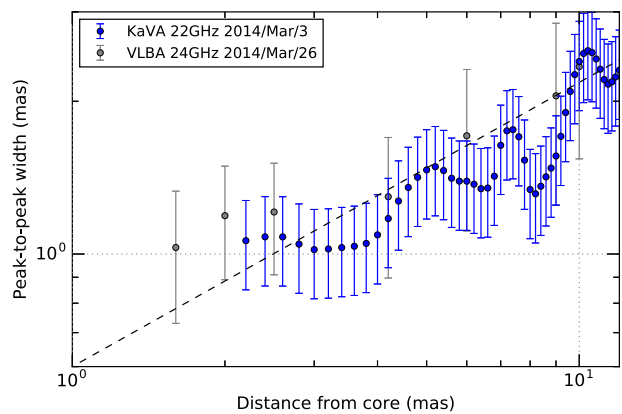}
 \end{center}
 \caption{Jet width profile of M87 as a function of distance from the core. Here the jet width is defined by the peak-to-peak separation of the north/south limbs, and measured every 0.2\,mas along the jet. The error bar on each KaVA data point is set to one-fifth of the width value. For comparison, we also plot the result from a contemporaneous VLBA 24\,GHz data, which was already presented in \citet{hada2016}. The dashed line represents $\propto r^{0.56}$ slope found by \citet{hada2013a} and \citet{asada2012} (here the line is not fitted to the data but is arbitrarily placed just for reference).}\label{fig:w-r}
\end{figure}

The uniformly weighted image (the bottom panel of Fig.~\ref{fig:m87map}) improves the angular resolution by $\sim$20\% ($\sim$1\,mas on average), resolving the rich structure of the inner jet regions. The well-known limb-brightened structure was clearly resolved up to $\sim$2\,mas near the core. With this image we measured the peak-to-peak width of the limb-brightened jet as a function of distance. The result is shown in Fig.~\ref{fig:w-r}, together with a contemporaneous VLBA 24\,GHz result presented in \citet{hada2016}. The KaVA result is in good agreement with the VLBA one in the overlapping regime, and the slope is also consistent with the previously known parabolic collimation profile. 

Note that M87 is known to have the counter jet (typically a-few-mas long) at the eastern side of the core~\citep{ly2007, kovalev2007}. For the present KaVA 22\,GHz data, we were also required to put some CLEAN components at the counter side through our deconvolution process, but the extent was limited only to $\lesssim$1\,mas. However. the contemporaneous higher-resolution VLBA images at 86/43/24\,GHz \citep{hada2016} found an exceptionally small extent of the counter jet around this period (visible only $\sim$0.5\,mas from the core). Thus the bare detection of the counter jet with KaVA is consistent with the VLBA results.

\subsection{Light curves}

\begin{figure}[ttt]
 \begin{center}
  \includegraphics[width=\columnwidth]{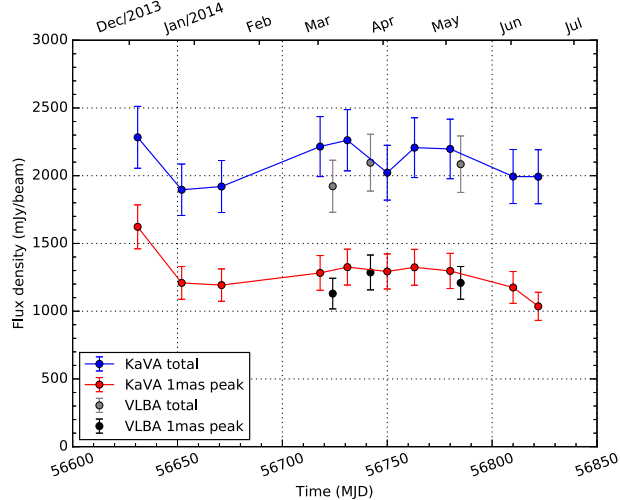}
 \end{center}
 \caption{M87 22\,GHz light curves observed during our program. The blue-colored light curve indicates the total CLEANed flux with KaVA, while the red-colored light curve represents the peak flux of the core when convolved with 1-mas-circular beam. For comparison, we also plot the results from our contemporaneous VLBA 24\,GHz data in grey (for the total) and black (for the peak) colors, respectively.}\label{fig:lc}
\end{figure}

Investigating a detailed radio light curve of the core or the innermost jet of M87 gives us important insight into high-energy activities near the black hole~\citep{acciari2009, hada2012, hada2014, akiyama2015, lee2016}. To check the feasibility of studying radio variability with KaVA, in Fig.~\ref{fig:lc} we show light curves for the M87 jet obtained through this program. Here we present two light curves describing different spatial scales: one is for the peak flux of the core when convolved with a 1-mas-circular beam, while the other is for the total CLEANed flux including the extended jet. In addition, we also plot available contemporaneous VLBA 24\,GHz light curves (the data presented in \cite{hada2016}) for the corresponding regions. For each data 10\% uncertainty is assumed in amplitude. 

For both the compact and the extended scales, the KaVA and VLBA flux densities are in good agreement with each other. For the core, the smooth flux variation is well sampled with KaVA especially for the period of continuous triweekly monitoring (2014 Mar--Jun). On the other hand, the measurement of integrated flux is generally more sensitive to the overall image sensitivity and uv-coverage. Nevertheless, thanks to the KVN baselines as short as 17\,M$\lambda$ (almost equivalent to the VLBA's shortest baselines 15\,M$\lambda$), KaVA consistently recovered the M87 extended emission (well within 10\% uncertainty) that was imaged by the VLBA. 

Therefore, these results demonstrate that the KaVA array is also very suitable for studying light curves of relativistic jets on various spatial scales.

\subsection{Structural evolution of the jet}
\begin{figure*}[htbp]
\begin{center}
\includegraphics[width = 0.95\textwidth]{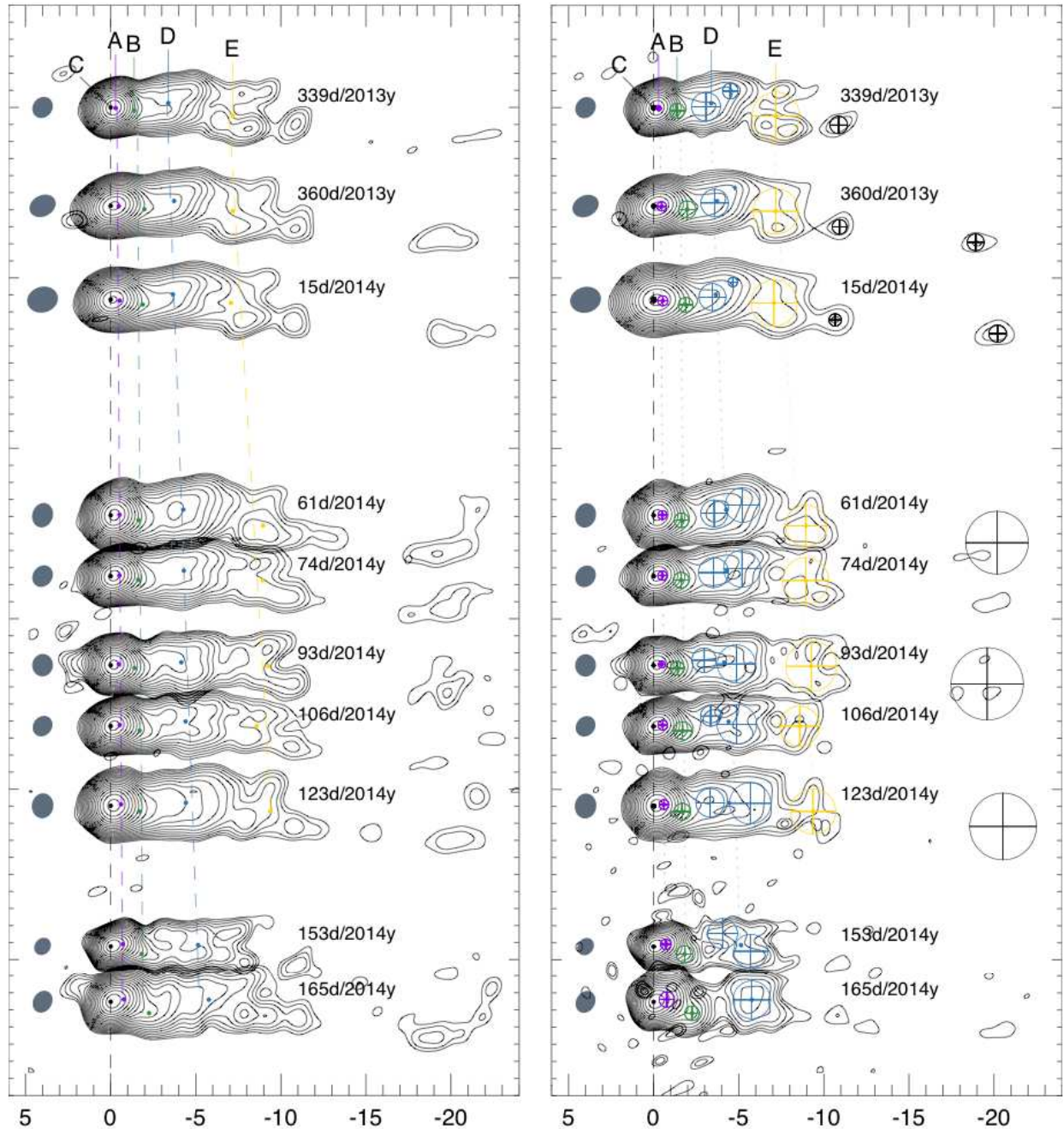}
\end{center}
\caption{\emph{Left} : Sequence of KaVA CLEAN images of M87 for all epochs. The model components, labelled A...E, obtained from {\tt modelfit} analysis are overlaid with coloured small circles. The vertical spacing is proportional to time elapsed. The dashed lines trace the best-fit linear motions of the components. The gray shaded ellipses at RA = 4 mas denote FWHM beam size. \emph{Right} : Images reconstructed by fitting multiple circular Gaussian models to visibility data. Component D is grouped by two Gaussian components at $\approx 3-4$ mas from core (see text for details). All the images are aligned with respect to the position of component C and are rotated by 20 degree in a clockwise direction. Contours start from 2.49 and 7.03 mJy / beam and increase by factors of $\sqrt{2}$ for the left and right figures, respectively. \label{evol}}
\end{figure*}

The structural evolution of the M87 jet is one of the major motivations of our monitoring program. On the left hand side of Fig.~\ref{evol}, we show multi-epoch CLEAN images of M87 between January and June 2014 (all the images are rotated by 20$^\circ$ clockwise). While more dedicated studies on the kinematics and structural evolution including two-dimensional velocity field are currently prepared in other papers (Park et al. in prep.; Ro et al. in prep.), here we focus only on the radial structural evolution of the jet based on a rather simple analysis, which can be quickly compared with various previous studies. 

Since the M87 jet has a limb-brightened and smooth intensity distribution,  rigorous component identification in the jet is non trivial~(e.g., \cite{mertens2016}). Nevertheless, here we applied the conventional visibility-based model-fitting technique (with {\tt modelfit} in Difmap), and reconstructed the M87 jet with a set of circular Gaussian components. These images are shown on the right hand side of Fig.~\ref{evol}. While this simple modeling often failed to recover the weak extended emission at $\sim$20\,mas, the inner jet within 10\,mas was reasonably characterized. This is supported by the fact that the number of fitted Gaussians and the derived parameters (location, size, flux) of each component in this region were remarkably similar among all the densely sampled images, indicating that the jet does have some ``characteristic patterns'' that can be commonly/consistently identified among different epochs. Of course the actual jet structure is much more complex than modeled here, but such a detailed modeling is beyond the scope of the present paper. A more complete modeling and characterization of this jet will be presented in a forthcoming paper (Park et al. in prep.)

\begin{figure}[!t]
\begin{center}
\includegraphics[width=1.0\columnwidth]{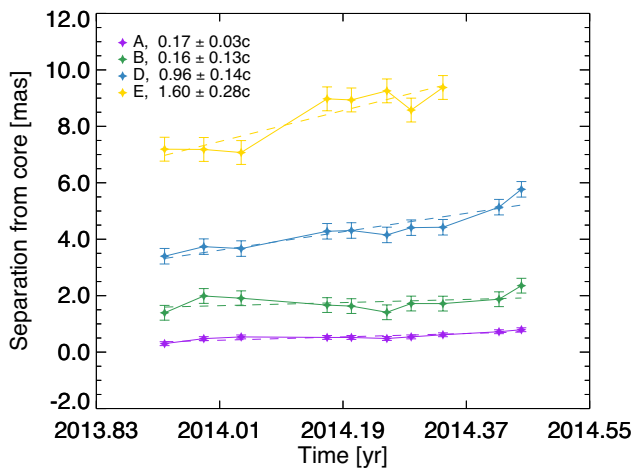}
\end{center}
\caption{Separation from core as function of time for various components (see the right panel of Figure~\ref{evol}). The dashed lines are the best fit lines and the obtained apparent velocities and their errors from the fitting are noted. We used an un-weighted fitting using the same errors for all the data points for each component that provide us with reduced $\chi^2$ of 1. \label{prop}}
\end{figure}

For each Gaussian component, the time evolution of the separation from the core is plotted in Fig.~5. Note that we grouped two Gaussian components at $\approx$3--5\,mas from the core and treated them as a single component (component D), estimating the flux-density-weighted component position. This grouping was made because (i) there is only a single Gaussian component in the last epoch at a similar location and (ii) their sizes and distribution vary a lot in different epochs but the grouped positions show a gradual outward motion and are smoothly connected to the single Gaussian component in the last epoch (see the left panel of Fig.~\ref{evol}). Thus the component D represents a spatially averaged motion around the 3--5\,mas region. The densely sampled dataset confirmed that all the components were moving outward. We derived an apparent velocity of each component by fitting a linear function to each dataset, and obtained a wide range of values from $\sim$0.16\,$c$ to $\sim$1.60\,$c$. These values are then plotted in Fig.~6, as a function of angular separation from the black hole, by using the result of the previous core-shift measurement by \citet{hada2011}\footnote{The result of \citet{hada2011} indicates that the convergence point of the core-shift is located at 70\,$\mu$as upstream of the 22\,GHz radio core. Here we assume that the convergence point is identical to the black hole location and the relative separation between the black hole and the 22\,GHz core is stationary with time.}. 

\begin{figure}[ttt]
\begin{center}
\includegraphics[width = 1.0\columnwidth]{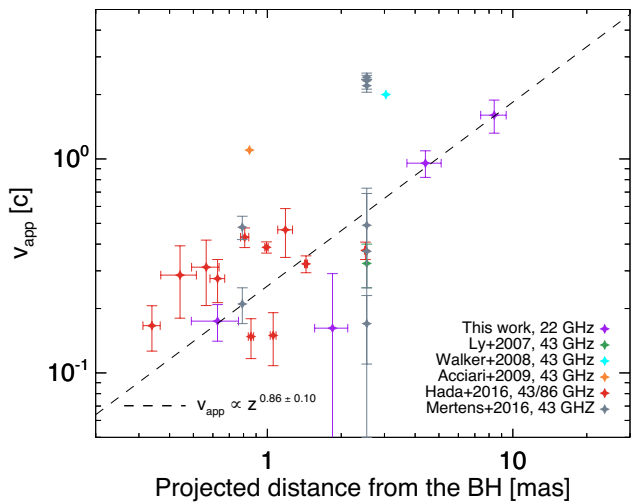}
\end{center}
\caption{Apparent velocities as function of projected angular distance from the black hole, obtained from {\tt modelfit} analysis (purple). The distance from the black hole was derived from the expected distance between the radio core at 22 GHz and the black hole of \citet{hada2011}. The values from the literature are shown with the data points with different colors. The error bars along the x-axis of our study and \citet{hada2016} show standard deviations of the fitted components positions. We note that we took a representative speed, $\approx 2c$, detected in the VLBA M87 movie project for \citet{walker2008} but a much denser distribution of apparent velocities obtained from the movie project has been provided in \citet{walker2016}. As for \citet{mertens2016}, we took the values from their stacked cross correlation analysis. The dashed line is the best-fit line to the data and the result of the fit is shown at the bottom left.}
\label{extended}
\end{figure}

As seen in Fig.~\ref{extended}, the apparent velocity profile obtained by KaVA suggests a trend of acceleration within 10\,mas from the black hole, showing a transition from sub-to-super-luminal motion around $\sim$5\,mas. For comparison, we also plot previous proper motion results at the same scales from the literature. The KaVA result is largely in agreement with those reported by \citet{ly2007}, \citet{walker2008}, \citet{mertens2016}, \citet{hada2016} and \citet{walker2016}, where they typically found slow motions at $\lesssim$1\,mas and fast motions beyond $\sim$1\,mas. As an additional test to check the validity of our measurement, we also applied the Wavelet-based Image Segmentation and Evaluation (WISE) technique \citep{mertens2015, mertens2016} to our KaVA data, and obtained a consistent velocity profile with that of our model fitting (see Appendix for more detail). This is additional support that our simple method is valid and the acceleration and superluminal motions obtained by KaVA are indeed real.

Note that, besides the moving components, our multi-epoch KaVA images suggest the presence of a quasi-stationary component at $\sim$20\,mas from the jet base. Interestingly, the M87 jet has a bright standing component at this distance since 1980's~\citep{reid1989, dodson2006}, implying a fundamentally different origin from the temporal moving features. We are investigating this component in more detail, and a complete analysis will be presented in a forthcoming paper (Park et al. in prep.). 

\begin{figure}[!t]
\begin{center}
\includegraphics[width = 0.95\columnwidth]{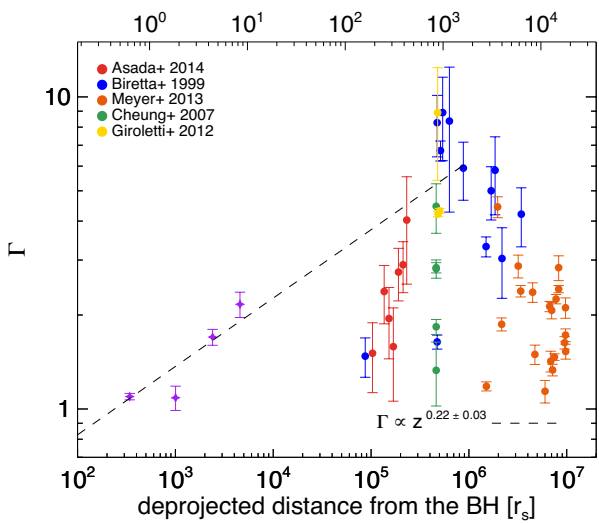}
\end{center}
\caption{Lorentz factor distribution as function of deprojected distance from the black hole in units of $R_{\rm s}$, which was derived from the assumed jet viewing angle of 15\,deg and the black hole mass of $6\times10^9M_{\odot}$ (purple diamonds). We present the values from the literature that probed Lorentz factors at large scales ($\gtrsim 10^5$\,$R_{\rm s}$) with circles with different colors.  The ticks on the top x-axis are projected angular distance from the black hole in units of mas.}
\label{extrapol}
\end{figure}

In Fig.~\ref{extrapol}, we show an extended plot of M87 jet kinematics including large-scale jet features studied in the literature~(\cite{biretta1999, cheung2007, giroletti2012, meyer2013, asada2014}). Adopting a viewing angle of 15 deg (e.g., \cite{perlman2011}) and a BH mass of $6\times 10^9 M_{\odot}$, the vertical and horizontal axes are described by Lorentz factor and deprojected distance from BH in $R_{\rm s}$ unit, respectively. The proper motion profile measured by KaVA ($v_{\rm app} \propto z^{0.86\pm0.10}$) is translated into $\Gamma \propto z^{0.22\pm0.03}$. Interestingly, this relation can consistently reproduce the maximum $\Gamma$ at HST-1 within $\approx 1\sigma$, implying that a single power-law acceleration profile may hold over three orders of magnitude in distance. Note that recent jet-shape studies of M87 at the same scales have independently discovered a collimation profile with a single power-law $z\propto r^{1.7}$~(\cite{asada2012, nakamura2013, hada2013a}). These results are consistent with the essence of the magnetically driven jet scenario, where the collimation and the acceleration are co-spatial~\citep{komissarov2009}. 


Of course the jet velocity profile obtained here is still rather coarse, and the exact profile may be more complex than suggested here, in fact as proposed by other detailed studies~\citep{asada2014, mertens2016}. Nevertheless, the important thing here is that this pilot program demonstrates that KaVA has excellent potential to study the velocity field and structural evolution of this jet in further detail. More extensive and advanced studies on this jet will be presented in the framework of the KaVA Large Program (as described below). 

\section{Summary and Future Prospects}
In this paper we have reported the basic performance of KaVA imaging and monitoring capability for M87 based on a pilot multi-epoch program at 22\,GHz. 
We have shown that KaVA 1\,Gbps observations of M87 at 22\,GHz can well image the extended jet structure down to $\sim$20\,mas from the core. An image dynamic range of $\sim$4000 was achieved, but further improvement of the KaVA imaging performance is expected depending on weather condition and integration time. The limb-brightened, parabolic jet structure at pc-to-subpc scales can also be imaged in agreement with VLBA images. This indicates that we can deeply study the morphology of this jet from a few 100\,$R_{\rm s}$ to $\sim$6000\,$R_{\rm s}$ with KaVA (for a viewing angle of 15$^{\circ}$). 

The densely sampled multi-epoch KaVA images have traced the detailed structural evolution of this jet. In agreement with the recent finding based on the VLBA 43\,GHz data by \citet{mertens2016}, our KaVA 22\,GHz data have also detected the superluminal outward motions together with linear acceleration within 10\,mas from the jet base. The independent confirmation at different time, at different frequency, and with different instruments suggests that the rapid jet acceleration near the jet base is not episodic but a fundamental property of the M87 jet. The fast jet speeds near the jet base may consistently explain why the counter jet of M87 is so weak. 

Finally, below we briefly describe the future prospects on our M87 program. 

\medskip

\noindent
\textbf{The KaVA Large Program:} While the kinematics result presented here is still rather coarse, this pilot study demonstrated that the KaVA has excellent potential to reveal the velocity profile of this jet in more detail. Following this, since 2016 we have upgraded our KaVA monitoring program of M87 in the framework of the KaVA Large Program~\citep{kino2015}. With the new program, every year we monitor M87 mostly biweekly over $>$6 months at both 22 and 43\,GHz quasi-simultaneously. Depending on the progress of our ongoing KaVA system upgrade, we will conduct observations at a higher recording rate ($\gtrsim$2\,Gbps). This strategy will greatly improve the quality of our proper motion analysis with KaVA and set a much stronger constraint on the velocity field of the M87 jet. More detailed studies about the jet structural evolution (both radially and transversally) are now actively ongoing by including the newer data. Also, the dual-frequency multi-epoch data will enable us to derive a set of accurate spectral index maps and their detailed evolution with time, which will provide important insights into the underlying particle energetics, density and magnetic fields. Furthermore, monitoring the light curve of the radio core on various time scales (from days to years) may tell us a hint on the accretion rate onto the black hole~\citep{park2017}, a key parameter to understand the jet production. 

\medskip
\noindent
\textbf{The East Asian VLBI Network:} Concurrently, the international VLBI collaboration in East Asia is rapidly growing. Besides the successful regular operation of the KVN and VERA array, the KaVA has started joint experiments with telescopes in China as well as some more stations in Korea and JVN. This will ultimately form the East Asian VLBI Network (EAVN), a huge VLBI array with up to $\sim$20 stations distributed throughout these countries~\citep{wajima2016}. The commissioning of EAVN is currently actively ongoing. In particular, in 2017, a part of our KaVA Large Program on M87 is expanded to invite some more stations: this includes the Tianma (Shanghai) 65\,m (22/43\,GHz), the Urumqi 25\,m (22\,GHz), the Nobeyama 45\,m (43\,GHz), the Takahagi 32\,m (22\,GHz), the Hitachi 32\,m (22\,GHz), the Kashima 34\,m (22\,GHz) and the Sejong 21\,m (22/43\,GHz). This will tremendously enhance the overall performance of our VLBI array such as angular resolution, sensitivity, uv-coverage and image dynamic range. With such a powerful capability of EAVN, we will be able to probe the acceleration and collimation properties of M87 over the whole distance from the jet base to HST-1 in unprecedented detail.   

\begin{ack}
We thank the anonymous referee for his/her careful review and suggestions that improved the manuscript. We acknowledge all staff members and students at KVN and VERA who supported the operation of the array and the correlation of the data. KVN is a facility operated at by the Korea Astronomy and Space Science Institute. VERA is a facility operated at National Astronomical Observatory of Japan in collaboration with associated universities in Japan. K.H. is supported by the Research Fellowship from the Japan Society of the Promotion of Science (JSPS). This work was partially supported by KAKENHI (26800109 and 15H00784). TJ and GYZ are supported by Korea Research Fellowship Program through the National Research Foundation of Korea (NRF) funded by the Ministry of Science, ICT and Future Planning (NRF-2015H1D3A1066561). I.C. is financially supported by the National Research Foundation of Korea (NRF) via Global PhD Fellowship Grant (NRF-2015H1A2A1033752). Part of this work was achieved using the grant of Visiting Scholar Program supported by the Research Coordination Committee, National Astronomical Observatory of Japan (NAOJ). 
\end{ack}

\section*{Appendix: Application of the WISE technique to KaVA data}

\begin{figure}[htbp]
\includegraphics[width = 1.0\columnwidth]{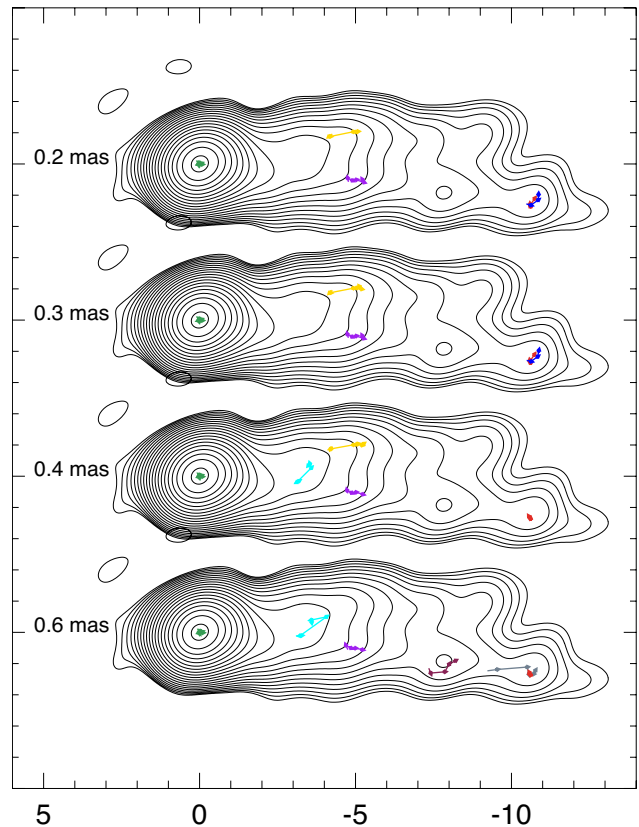}
\caption{Arrows showing the observed displacements of the significant structural patterns (SSPs, see  \cite{mertens2015, mertens2016} for more details) using the WISE technique overlaid on contours of the stacked CLEAN images of all 10 epochs data rotated by 20 deg in a clockwise direction. The results of using different spatial decomposition scales, 0.2, 0.3, 0.4, and 0.6 mas, are shown. \label{wise}}
\end{figure}

\citet{mertens2016} showed that the detailed velocity field of the M87 jet can be investigated using the Wavelet-based Image Segmentation and Evaluation (WISE) method. This method provides an \emph{un-supervised} scheme to detect and identify various characteristic structural patterns in different epochs and measure the jet velocities. We applied the WISE method to our data set and obtained velocity vectors at different regions of the jet on four spatial (decomposition) scales, 0.2, 0.3, 0.4, and 0.6 mas (Fig.~\ref{wise}, see e.g., \cite{mertens2015} for details)\footnote{We followed the command line tutorials for 3C 120 provided on the Web site; http://flomertens.github.io/wise/tutorials\_cmd/Walkthrough3C120.html}. Most of the velocity fields showed radial, outward motions except quasi-stationary ones at $\approx 0$ and $\approx 10$ mas from the core on all the four scales (green and red arrows, respectively), and another one at $\approx 3$ mas from the core on a scale of 0.6 mas (cyan arrow) showing back-and-forth motions. This behavior may reflect the complex nature in the 3--5\,mas regions, as we needed a component grouping in this area in our modelfitting approach (see Section 3.3). In addition, we detected at most four to six velocity vectors only at specific locations in the jet. These numbers are quite smaller than those presented in \citet{mertens2016}. These are likely due to the limited angular resolution of KaVA, not to the limited sampling of our monitoring since the average interval between adjacent epochs is $\approx$ 19 days, which is comparable to that of \citet{mertens2016}.

\begin{figure}[htbp]
\includegraphics[width = 1.0\columnwidth]{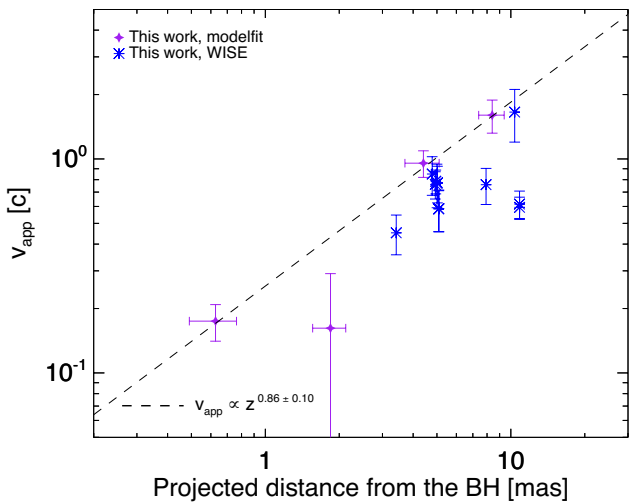}
\caption{Same as Fig.~6 but excluding the values from the literature and including the apparent velocities of the radial, outward motions from the WISE technique with blue asterisks (see text for details). \label{velwise}}
\end{figure}

In Fig.~\ref{velwise} we present the apparent velocities of the radial, outward motions derived from the WISE technique. Note that not all these values are independent of each other. The values obtained from the WISE method show a consistent acceleration trend with our modelfit analysis, although they seem to be slightly below our best-fit line. This small difference might be due to an under-estimation of the velocities in the WISE method or an over-estimation in our modelfit analysis. This could also be investigated in our future studies using the Large Program data in more detail (e.g., Park et al. in prep.).

\end{document}